# An Elementary Derivation of Mean Wait Time in Polling Systems

Field Cady
Think Big Analytics LLC

## Abstract

Polling systems are a well-established subject in queueing theory.  However, their formal treatments generally rely heavily on relatively sophisticated theoretical tools, such as moment generating functions and Laplace transforms, and solutions often require the solution of large systems of equations.  We show that, if you are willing to only have the average waiting of a system time rather than higher moments, it can found through an elementary derivation based only on algebra and some well-known properties of Poisson processes.  Our result is simple enough to be easily used in real-world applications, and the simplicity of our derivation makes it ideal for pedagogical purposes.

## Introduction

Polling systems are a classic subject in stochastic analysis, applicable to such diverse areas as computer hardware and elevator performance.  In a polling system there are N queues which receive jobs, and a single server processes all queues.  The server visits the queues in a deterministic order, stopping at each queue to process any jobs which might be there, before continuing along its path.  Depending on the specific system, when the processor arrives at a queue it may process jobs until the queue is empty (usually called "exhaustive processing") or process only those jobs that are in the queue at the time of arrival ("gated polling").

Polling systems have been treated by many authors, both as a subject in themselves and as models for applied situations.  Some of the earliest works were in the context of modeling the performance of hard disks, where each track on the disk is seen as a distinct queue and the jobs are read/write memory requests.  Several of the earliest derivations [1, 2] were later found to contain subtle errors, which were eventually corrected at the cost of more complicated derivations and more unwieldy final results.  Typically in the literature, determining quantities of interest for a polling system (such as average queue length or waiting times at the different queues) requires numerically solving a set of K equations, where K is polynomial in the number of queues [3-5].  This complexity makes these analytical results impractical for many real-world situations.  For example, the effort required to understand a derivation, formulate the full set of equations for a particular system, and solve them numerically is often less than the effort required to exhaustively simulate the system and evaluate its performance empirically.

[6] is one of the few works that trades off fine-grained results for a simple derivation and easy-to-use formula. They studied the performance of hard disks by modeling them as a polling system with a continuum of queues. Their derivation uses no math beyond basic calculus, and they give a closed formula for the system's waiting time rather than a set of equations (on the other hand, their approach does not give higher moment of waiting time, or waiting times for the different queues). The current paper generalizes that work to discrete queues, and discusses it in the more general context of polling systems.

Our analysis is applicable to any order in which the queues are served, and unlike previous work it allows for the distribution of job size to vary by queue (though it requires that the average job size be the same across all queues). The key to the proof is an accounting technique we call "shuffling" (the terminology used by [6]), which breaks the mean wait time into two easy-to-calculate terms.

In the next section we give an overview of our model and the terminology we use; the model is deliberately simple for clarity of exposition. Next we present the derivation of mean wait time, with an eye toward how essentially the same derivation (perhaps with a bit more calculation) can be applied to related models. Finally, we discuss some applications of our work and directions for future research.

## Model

We assume a system of N queues and a processor which moves between them in round-robin order (1, 2, …, N-1, N, 1, 2, …). Jobs arrive as independent Poisson processes to each of the N queues with rates $\lambda_1, \lambda_2, \ldots \lambda_N$. The sizes of the jobs arriving at queue i are i.i.d, and we use the random variable $S^i$ to denote the time required to complete a job at the queue i. We do not require that the $S^i$ be identically distributed, but we do require that their mean sizes be the same. We use the random variable $S$ to denote the size of a job arriving into the system regardless of its queue, and we denote the overall arrival rate of jobs by $\lambda = \lambda_1 + \lambda_2 + \ldots + \lambda_N$. We let $\rho = \lambda E[S]$ denote the total load on the system, and we assume that $\rho < 1$, so that the system is not overloaded and a steady-state distribution can be assumed to exist.

We assume that the processor takes time $\alpha$ to move from one queue to the next. Processing at a given queue is exhaustive; the processor will work until all jobs are finished, including any that may have arrived while it was working.

We define the random variable W to be the waiting time between when a job enters the system and when the server begins work on it. The objective of this paper is to calculate E[W].

We have made a number of simplifying assumptions which make our model simple and our derivation less messy, but which are not actually required. We have assumed circular polling, but in fact the queues could be visited in any order that is oblivious to the location of jobs in the system; for example, it would cover elevator scanning, or even a situation where the next queue the server visits is picked at random. We assume a constant switching time $\alpha$, but the times to go between pairs of queues could be of different, or even random values. We assume that the jobs at a queues are processed in FCFS order, but in fact any size-oblivious order would work as well (such as LCFS). Finally, we have assumed the processor is exhaustive, but gated processing is equally well covered. We will point out where our derivation can be modified to accommodate these changes.

# Derivation

There are two key ideas to our derivation. The first is a way to break E[W] into two parts, one capturing the time spent waiting while the server is working on other jobs, and the other capturing the time spent waiting while the server is moving between queues. The first of these can be solved using an accounting method we call "shuffling" (the second key idea in our proof), and the second is straightforward to calculate.

We define the following random variables:

$$M = portion\ of\ W\ during\ which\ the\ server\ is\ moving$$
$$P = portion\ of\ W\ during\ which\ the\ server\ is\ working\ on\ a\ job.\ Equals\ W - M$$

It is clear that W=P+M, and hence

$$E[W] = E[P] + E[M]$$

We will calculate E[M] and E[P] separately and combine them to find E[W].

## Calculating E[P]

We calculate E[P] using the technique of "shuffling" described by Cady et al. The key heuristic for understanding shuffling is that whenever a job enters the system it arrives with a label attached to it, and when the server begins working on the job, the label that is attached to it leaves the system. The catch though is that labels waiting within the system can be permuted in any way and at any time, so when a job begins processing it may be attached to a different label from the one it arrived with. Focusing on labels in lieu of the jobs themselves allows us to de-couple the time a job arrives in the system from the time it begins getting processed.

Because shuffling the labels doesn't affect the actual jobs, the behavior of the system is not affected by shuffling: shuffling is merely an accounting technique. And critically, there are always the same number of labels in the system as there are jobs waiting. The labels may not be the ones that originally arrived with the jobs, but the total number will be the same at any point in time. It is also clear that we can define the random variable P analogously for labels; the portion of the time a *label* spends in the system during which the server is processing a job. If the system goes on for a very long time T and processes K jobs/labels, it is clear that

$$E[P] = \frac{1}{K}\sum_{i=1}^{K} P_i = \frac{1}{K}\int_0^T N(t) * 1\{\text{server is working on a job}\} dt$$

where N(t) is the number of jobs/labels in the system at time t. Because shuffling does not affect N(t) and does not change when the server is working on a job, we can see that E[P] for labels is the same as E[P] for jobs, and that it is invariant under shuffling (although higher moments of P for labels may not be invariant). So in calculating E[P] we may focus on labels instead of jobs and pick a convenient shuffling protocol.

The calculation is made very convenient if we force the labels to be processed in FCFS order. That is, in the instant before the server begins work on its next job, shuffle the labels so that the label attached the next job is the label that has been waiting in the system the longest. For calculating E[P] we pick an arbitrary label L which arrives at time t, and denote its P by $P^L$. There are two components of $P^L$:
1. The time to finish the job being processed at time t, if there is one
2. The time to wait for all labels that are in the system when L arrives

Term (1) is affected by the fact that larger jobs take longer to be processed, and hence the distribution of job size for a job being processed is different from that of all jobs; this is often called the "inspection paradox". It is a standard result that the average remaining time on a job being processed is $E[S^2]/2E[S]$, meaning that the average value of (1) is $\rho E[S^2]/2E[S]$ (since there will be such a job a fraction $\rho$ of the time).

Calculating (2) is more subtle. We know that there are N(t) jobs/labels in the system when L arrives, and L will have to wait until these N(t) labels have been processed. In general there may be a very complicated probabilistic relationship between N(t), how the N(t) jobs are distributed among the queues, and how the next N(t) jobs that actually get processed will be distributed among the queues (remember that the next N(t) jobs that get processed may include some that haven't arrived yet). However, we have made the critical assumption that $E[S_i]$ is the same for every i, and so the average value of (2), conditioned on N(t), will be N(t)E[S].

By the PASTA property of Poisson processes ("Poisson arrivals see time averages"), we know that the average number of labels waiting in the system when L arrives will be the time-average of N(t). We know that the throughput of labels is $\lambda$, and the average amount of time a label spends in the system is E[W]. So Little's Law tells us that
$$E[N(t)] = \lambda E[W]$$
and hence that
$$E[P] = \rho E[S^2]/2E[S] + (\lambda E[W])E[S] = \rho E[S^2]/2E[S] + \rho E[W]$$

Note that our derivation of E[P] has only relied on the Poisson arrival of jobs, the fact that queues process them in a size-oblivious order, and the fact that the average job size is the same for all queues. It applies equally well to any size-oblivious scheduling policy at the queues (such as LCFS) and any protocol for moving the server.

# Calculating E[M]

When a particular job arrives there are two possible states that the processor can be in :
1. It is moving between two queues, which has probability $1-\rho$
2. It is at a queue processing jobs, which has probability $\rho$

In the first case the distribution of the location of the processor will be uniformly distributed over its path, meaning the average time will be $N\alpha/2$.

For the second case we imagine the job arrives at queue j when the processor is at queue i. We define

$$\pi_{ij} = \text{time to move from queue } i \text{ to queue } j$$

In our case of circular polling, fixed switching times between queues, and exhaustive processing, we have

$$\pi_{ij} = \alpha(j-i) \quad \text{if } j>i$$
$$= \alpha(N-j+i) \quad \text{if } i<j$$
$$= 0 \quad \text{if } i=j$$

We know that the probability of the processor being at queue i when the job arrives is $\rho_i$ and the probability that the job will arrive in queue j is $\rho_j/\rho$. Summing over the possible values of i and j, and combining terms (1) and (2) above, we see that

$$E[M] = (1-\rho)\frac{N\alpha}{2} + \frac{1}{\rho}\sum_{i,j}\rho_i\rho_j\pi_{ij}$$

In our case of circular polling, fixed switching times between queues, and exhaustive processing, this becomes

$$E[M] = (1-\rho)\frac{N\alpha}{2} + \frac{1}{\rho}\sum_{i,j}\rho_i\rho_j\pi_{ij}$$
$$= (1-\rho)\frac{N\alpha}{2} + \frac{1}{\rho}\sum_{i<j}\rho_i\rho_j\alpha(j-i+N-j+1) + \sum_i 0$$
$$= (1-\rho)\frac{N\alpha}{2} + N\alpha\frac{1}{\rho}\sum_{i<j}\rho_i\rho_j$$
$$= (1-\rho)\frac{N\alpha}{2} + N\alpha\frac{1}{2}(\rho - \frac{1}{\rho}\sum_i\rho_i^2)$$
$$= \frac{N\alpha}{2}\left\{1 - \rho\sum_i(\rho_i/\rho)^2\right\}$$

For simplicity we will denote this expression by $\Pi$. Note that $\rho\sum_i(\rho_i/\rho)^2$ is the probability that 1) L arrives when the server is working on a job, and 2) L arrives in the queue the server is at. So $\Pi$ is really just half the cycle time of the server times the probability that it has to move at all before L gets processed.

Gated processing by each queue could be handled by simply setting $\pi_{ii}$ to $N\alpha$ instead of 0; you could even allow some queues to be gated and others exhaustive. For more complicated polling orders, like an elevator scan which changes directions, you can use i to index not just the queue that the processor is at, but the state that it is in (for example, moving up from 1 toward N versus moving down).

### Combining the results

Combining the expressions for E[J] and E[M] we get

$$E[W] = E[S^2]/2E[S] + \rho E[W] + \Pi$$
$$E[W] = \frac{\rho}{1-\rho} E[S^2]/2E[S] + \frac{1}{(1-\rho)}\Pi$$

## Discussion

This paper does not present a fundamentally new result; in fact, the result is weaker than that in previous works. Its strength lies instead in the simplicity of the derivation and of the resulting formula. This makes it much more practical in applied situations where ease-of-use is more important than detailed information about the system. In fact, many real-world systems do not satisfy simplifying assumptions like the Poisson arrival of jobs, and so simulation is a necessary step in creating applications. In these cases, simple derivations like ours can be used to guide initial work prior to exhaustive simulation. They can also provide a basis for understanding how much non-ideal properties of the system actually impact the bottom line.

This paper is also potentially important for the way it breaks E[W] down and for the technique of shuffling. These ideas can be used in a range of situations where it is sufficient to know only the first moment of a system's waiting time.

Finally, the complexity of polling system derivations in the literature, and their usual reliance on fairly sophisticated mathematical tools, puts them out of reach of many students. On the other hand our derivation requires no high-level math and uses only a minimum of calculation, making it a suitable topic for introductory classes in stochastic analysis. In addition, the final formula is elegant and the derivation contains non-trivial conceptual steps, which are important for making queueing theory appealing to students while also teaching them to think critically.

## References

[1] E. G. Coffman, L. A. Klimko, and B. Ryan, "Analysis of scanning policies for reducing disk seek times," SIAM Journal on Computing, vol. 1, no. 3, pp. 269–279, 1972.

[2] W. C. Oney, "Queueing analysis of the scan policy for moving-head disks," Journal of the Association for Computing Machinery, vol. 22, pp. 397–412, 1975

[3] Takagi, H. (1988). Queuing analysis of polling models. *ACM Computing Surveys (CSUR)*, *20*(1), 5-28.

[4] Winands, E. M., Adan, I. J., & Van Houtum, G. J. (2006). Mean value analysis for polling systems. *Queueing Systems*, *54*(1), 35-44.

[5] Winands, E. M. M., Adan, I. J. B. F., & van Houtum, G. J. (2007) MVA for polling systems: an efficient approach. *Analysis of Manufacturing Systems.*

[6] Cady, F., Zhuang, Y., & Harchol-Balter, M. (2011). A Stochastic Analysis of Hard Disk Drives. *International Journal of Stochastic Analysis*, *2011*.